\documentstyle[11pt,paspconf,psfig,rotate]{article}
\begin{document}
\title{Environments of QSOs at intermediate redshifts}
\author{K.J\"ager and K.J.Fricke}
\affil{Universit\"ats-Sternwarte G\"ottingen, Germany}
\author{J.~Heidt}
\affil{Landessternwarte Heidelberg, Germany}
\begin{abstract}
A preliminary analysis of fields around 20 mainly radio--quiet 
QSOs (RQQs) at intermediate redshift is summarized. We find overdensities 
of faint sources around 50\% of our observed QSOs suggesting that they 
are located in groups or even clusters of galaxies. 
\end{abstract}
\section{Introduction}
AGN environment studies, from host galaxy to galaxy cluster scales, provide 
information on the link between AGN activity and tidal interactions and of 
differences of the environments due to intrinsic AGN properties or an 
evolution with z. Obtained results have to be considered in particular 
within unified schemes of active galaxies. Following earlier studies at 
low z, RQQs reside in
spiral hosts and within poor environments while radio--loud QSOs (RLQs) 
reside in elliptical galaxies and tend to lie within clusters of Abell richness 
class 0--1. But this picture has gained more complexity since some RQQs have 
recently been found also within elliptical hosts (e.g.~Bahcall et al.~1997). 
Moreover, few observations at z$>$1 seem to show both RLQs and RQQs residing 
in comparable compact groups of possibly starbursting galaxies (e.g.~Hutchings 
et al.~1995) which would suggest a change of the environments of RQQs at 
intermediate z. 
\section{Observations, measurements and results}
During an ongoing multicolour imaging survey of QSOs, we obtained deep R band 
images (R$\approx$24.5, 0.8''$<$FWHM$<$1.7'') of 20 QSOs within the range 
0.75$<$z$<$0.85. Observations were made with the Calar Alto-- and the 
ESO/MPG--2.2m telescopes using CAFOS and EFOSC2 and total integration 
times of typically $\approx$1h$/$target of the finally coadded frames.
The subsample presented here contains 16 RQQs (no radio flux in the
Veron--catalogue) and 4 RLQs. All fields were carefully selected 
using the DSS and NED to avoid e.g. the contamination by known galaxy clusters.  
For detection, deblending, classification and photometry of sources 
we used the SExtractor package (Bertin \& Arnouts 1996) accompanied by manual 
checks of the
frames. In almost all cases, faint close companion candidates of the QSOs, 
or even hints for tidal interactions are present, suggesting that the QSOs are 
not isolated. In figure 1 (right) we show e.g.~the 20''(110kpc at z$_{QSO}$,
H$_{0}$=75, q$_{0}=$0.5) field around 1211+0848 (z=0.81, center) 
including a lightbridge from the QSO towards a faint galaxy.
As a first object distribution analysis of all fields, we measured the number 
of faint (R$>$20) sources within boxes of increasing size 
(r$=$100--500kpc), each centered on the QSOs. Then we compared these number 
densities with the average object density obtained from the field outside the 
largest box. Due to our large field sizes ($>$8'), an appropriate 
determination of field counts could be obtained. Figure 1 (left) shows e.g.~an 
increase of the number counts towards Q 1340--0020 (z$=$0.79) compared to the 
average counts, which are represented by the short dashed line (and their 
Poission scatter represented by the long dashed lines). Error bars of the box 
counts are also from Poission statistics. 
In 50\% of all cases we have found a trend of increasing number 
counts towards the QSOs (8 of 16 RQQs, 2 of 4 RLQs), whereas in five
cases an excess appears more compact and is concentrated within $<$200kpc 
around the QSOs. We suggest that they reside at least in groups of galaxies.
This preliminary analysis already shows evidence for RQQs at
intermediate z to be located in a diversity of sourroundings more 
comparable to RLQs than to their low redshift counterparts.
Our further detailed statistical and multicolour analysis as well as an 
spectroscopic follow--up program with the VLT and HET should confirm,      
if we either see physical associations of galaxies with the QSOs or foreground
objects.
\begin{figure}
{\centerline{{\rotate[r]{\psfig{figure=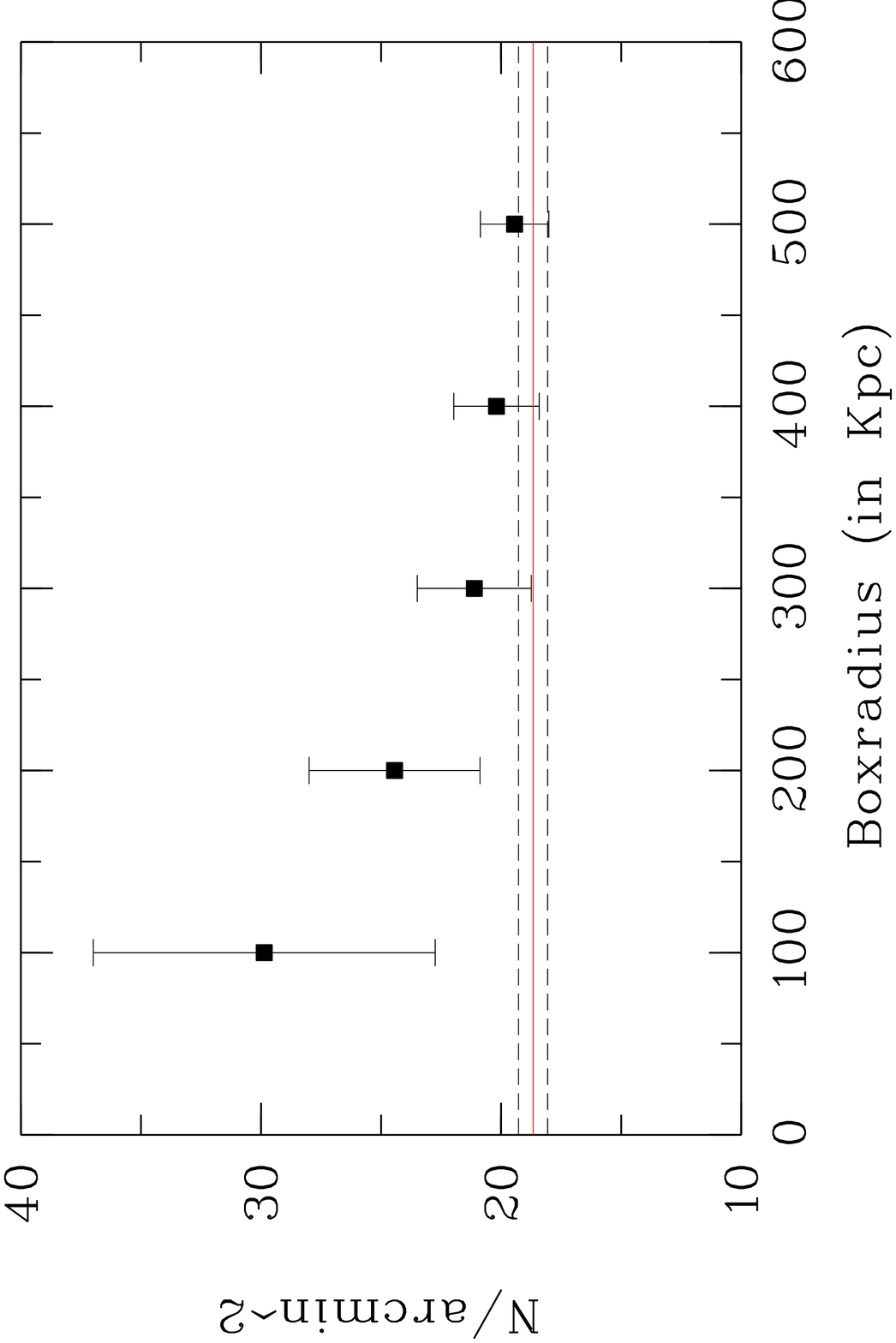,height=6cm,width=5.2cm}}}
{\psfig{figure=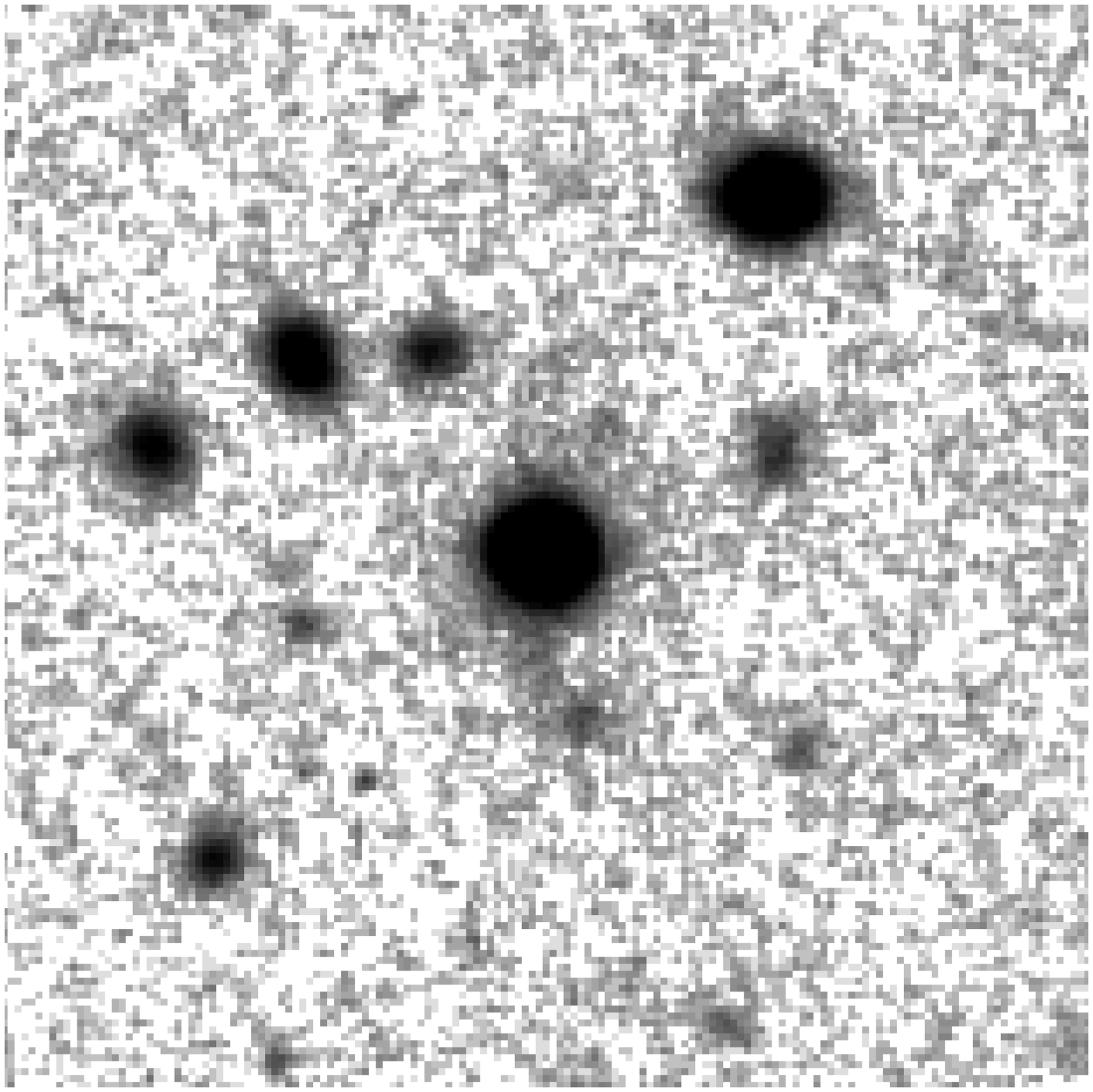,height=5cm,width=5cm}}}}
\caption{Environment statistics of QSO 1340--0020 (left) and the close 
environment of QSO 1211+0848 (right), cf.~text.} \label{fig-1}
\vspace*{5mm}
\end{figure}
\acknowledgments
This work has been supported by the Deutsche \\
Forschungsgemeinschaft (DFG--grant 
FR 325/41--1).

\end{document}